# A publicly available vessel segmentation algorithm for SLO images

Adam Threlfall,*[1,2] Samuel Gibbon,*[1,2] James Cameron,[3] Tom MacGillivray[1,2]

*Shared first authorship

[1] Centre for Clinical Brain Sciences, University of Edinburgh, UK

[2] Robert O Curle Ophthalmology Suite, Institute for Regeneration and Repair, University of Edinburgh, UK

[3] Princess Alexandra Eye Pavilion, NHS Lothian, Edinburgh, UK

**Abstract**

**Background and Objective:** Infra-red scanning laser ophthalmoscope (IRSLO) images are akin to colour fundus photographs in displaying the posterior pole and retinal vasculature fine detail. While there are many trained networks readily available for retinal vessel segmentation in colour fundus photographs, none cater to IRSLO images. Accordingly, we aimed to develop (and release as open source) a vessel segmentation algorithm tailored specifically to IRSLO images.

**Materials and Methods:** We used 23 expertly annotated IRSLO images from the RAVIR dataset, combined with 7 additional images annotated in-house. We trained a U-Net (convolutional neural network) to label pixels as "vessel" or "background".

**Results:** On an unseen test set (4 images), our model achieved an AUC of 0.981, and an AUPRC of 0.815. Upon thresholding, it achieved a sensitivity of 0.844, a specificity of 0.983, and an F1 score of 0.857.

**Conclusion:** We have made our automatic segmentation algorithm publicly available and easy to use. Researchers can use the generated vessel maps to compute metrics such as fractal dimension and vessel density.

**Introduction**

Infra-Red Scanning Laser Ophthalmoscope (IRSLO or SLO) images, captured by devices such as the *Heidelberg SPECTRALIS* (Heidelberg Engineering, Heidelberg, Germany) and *i-Optics EasyScan* (EasyScan, The Netherlands), are rich in the detail of the fundus that they capture. However, they are underutilised for tasks such as retinal vascular analysis, despite yielding high-resolution images with excellent contrast that could offer advantages over conventional colour fundus photography (see Figure 1). In previous work, we adapted the Vascular Assessment and Measurement Platform for Images of the REtina (VAMPIRE) software[1] (which at the time used traditional image processing techniques to segment the vessels in colour fundus photographs), to accept IRLSO images,[2] demonstrating high reliability. More recently, deep learning has surpassed the performance of such traditional techniques, and has excelled in vessel segmentation for colour fundus photography[3–5] and various SLO modalities such as pseudo-colour SLO[6–9] and ultra-widefield SLO.[10] However, few researchers have focused on IRSLO image segmentation. Existing attempts, such as those by Ajaz *et al.*[11] and González Godoy,[12] have yielded promising results, however the trained models are not publicly available. In this work, we introduce a deep learning model for SLO image vessel segmentation and make it publicly available (https://github.com/adamthrelfall/IRSLO-Segmentation).

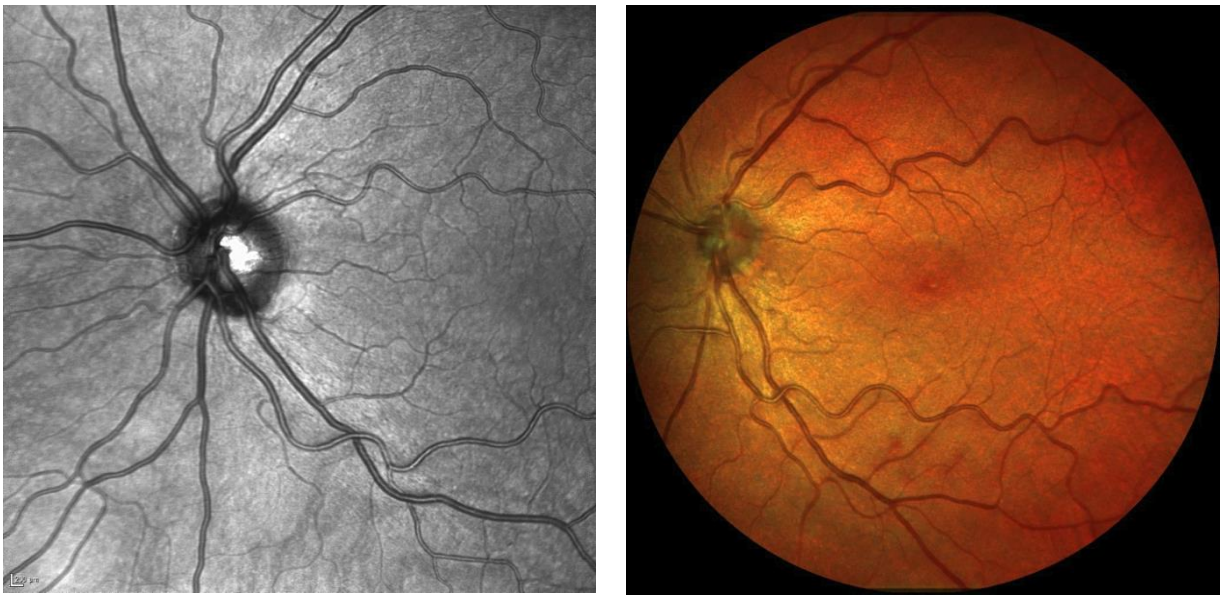

**Figure 1:** Scanning Laser Ophthalmoscope (SLO) images from two different devices. Heidelberg Spectralis (left), i-Optics EasyScan (right). Images taken from the RAVIR[13,14] and IOSTAR[6,7] datasets respectively.

# Methods

We leveraged the commonly-used UNet deep learning architecture[15] for the task of segmenting SLO images into two classes: vessels and background tissue. This network was chosen because it (and its derivatives[16–18]) have previously demonstrated excellent performance in a range of medical image processing tasks including vessel segmentation in colour fundus photography.[4,19]

We trained the network on a total of 30 images from two sources: the publicly available RAVIR dataset of 23 annotated images (resolution 768 × 768 pixels),[13,14] and 7 images from our previous work (resolution 1536 × 1536 pixels),[2] manually annotated in-house (author SG) using ITK-Snap (http://www.itksnap.org/pmwiki/pmwiki.php), which supports pixel level annotations. The RAVIR dataset includes vessels labelled as arteries and veins, however for the current work we combined these into a single "vessel" class. All images were captured with a *SPECTRALIS* device. They were split into training, validation, and test sets of 24, 2 and 4 images respectively. Each image was split into 20 randomly chosen windows of 320 × 240 pixels. This approach was implemented for three key reasons: Firstly, individual data augmentation was applied to each window, enhancing the diversity of the training dataset. Secondly, it effectively reduced computational overhead by consuming less memory for each window than would be consumed for a full-sized image. Thirdly, it enabled multiple instances of backpropagation per eye, thereby accelerating the training rate.

For training, we used the Adam optimiser and Dice Focal loss, and performed training over 600 epochs (300 with a learning rate of $1 \times 10^{-3}$, and 300 with a learning rate of $1 \times 10^{-4}$), with a batch size of 20. These parameters were empirically determined to provide sufficient time for training (therefore avoiding underfitting), whilst minimising training time.

Data augmentation was performed to increase generalisability of the dataset. The augmentation included random affine transformations (resizing, rotation, and shear), random intensity operations (histogram equalisation, contrast-limited adaptive histogram equalisation, intensity rescaling, and logarithmic intensity operations), and random gaussian blur. Not all operations were performed on each image, rather, a random subset of augmentation operations were performed at training time to ensure that, during each epoch, the network was learning from slightly different training data, therefore reducing overfitting.

# Results

The results for the network are shown in Table The receiver-operator characteristic (ROC) and precision-recall curve (PRC) are shown in Figure 2.

**Table 1:** Accuracy of the trained networks on an unseen test set.

| Parameter | Result |
| --- | --- |
| AUC | 0.981 |
| AUPRC | 0.815 |
| Sensitivity | 0.844 |
| Specificity | 0.983 |
| F1 Score | 0.857 |
| Accuracy | 0.966 |

**Abbreviations:** AUC (area under the receiver operating characteristic curve), AUPRC (area under the precision-recall curve).

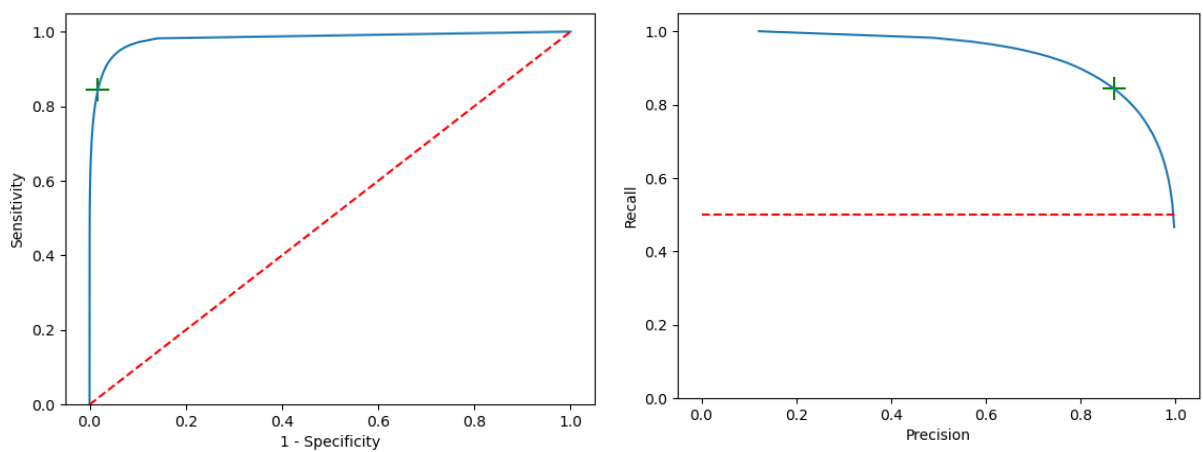

**Figure 2:** Receiver-operator characteristic (left) and precision-recall curve (right) for network segmentation. The green cross indicates the threshold point which optimises F1 score (threshold = 0.45, F1 = 0.857).

On a test system using an Nvidia Quadro RTX 5000 GPU and Intel i9 10900X CPU with 32GB of RAM, running Windows 11, the segmentation took 0.03 seconds for a 768 × 768-pixel image, and 0.13 seconds on a 1536 × 1536-pixel image. This speed means that this network could be applied in a clinical setting for near-instant segmentation of captured images. Using the CPU only, segmentation takes approximately 0.8 seconds for a smaller image, or 3.6 seconds for a larger image. On a less powerful Intel i3-1115G4 laptop CPU, 8GB of RAM and without GPU acceleration, segmentation takes approximately 80 seconds for a 1536 × 1536-pixel image. Output vessel maps are demonstrated in Figure 3.

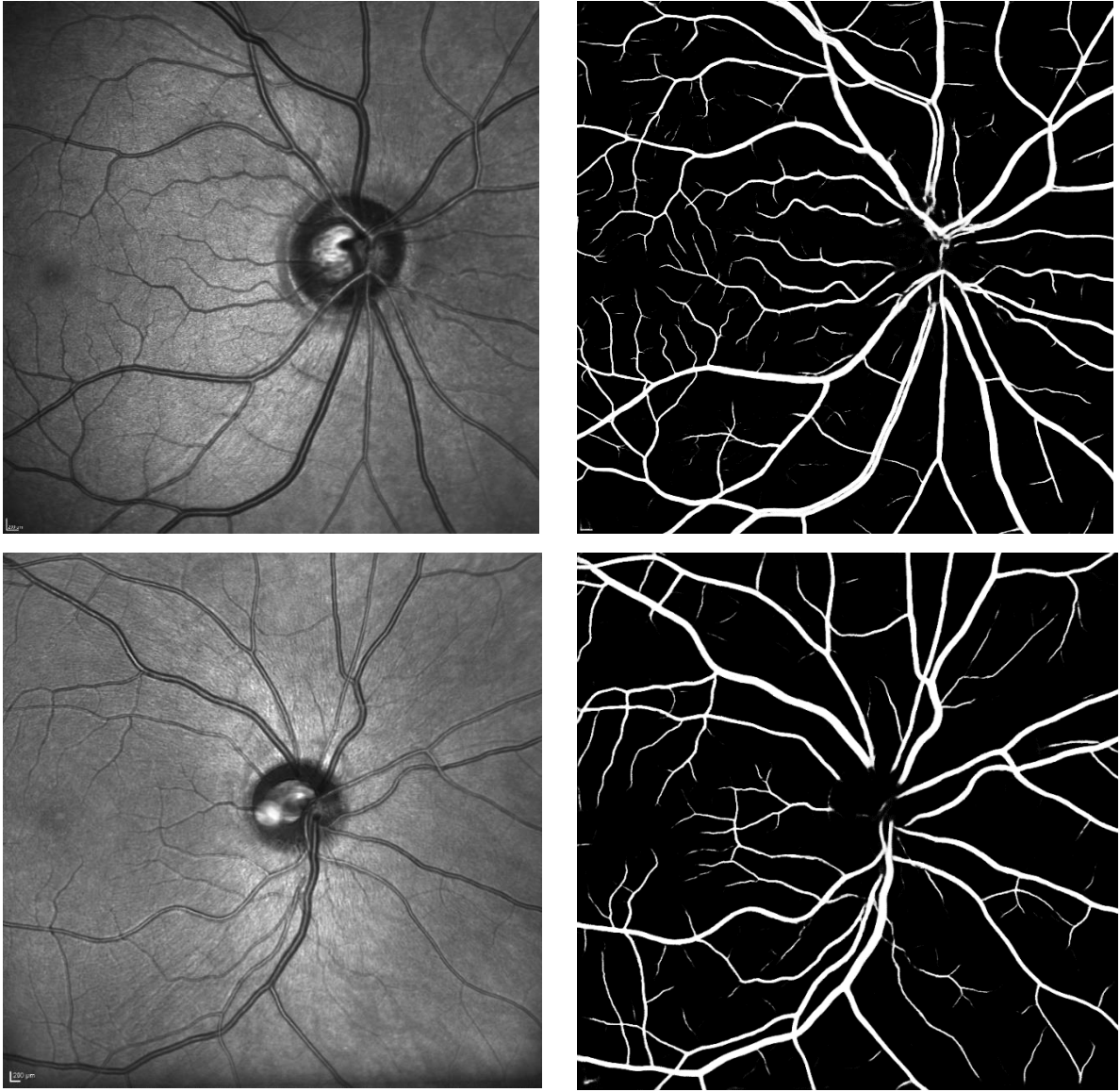

**Figure 3:** Original SLO images (left) and automatically generated vessel maps (right). The top image has a resolution of 1536 × 1536 pixels, the bottom has a resolution of 768 × 768 pixels.

## Discussion

We have demonstrated that a UNet accurately and quickly segments the vasculature appearing in IRSLO images. Our method is seen to reliably segment vessels across a range of features (e.g. retinal lesions, non-uniform illumination or poor focus) due to the presence of these features in the RAVIR dataset.[13,14]

An example segmentation of an atypical image is provided in Figure 4, demonstrating how data augmentation makes the network more robust to the tortuous vessels, local blur, and non-uniform illumination which is often seen in real-world patient images.

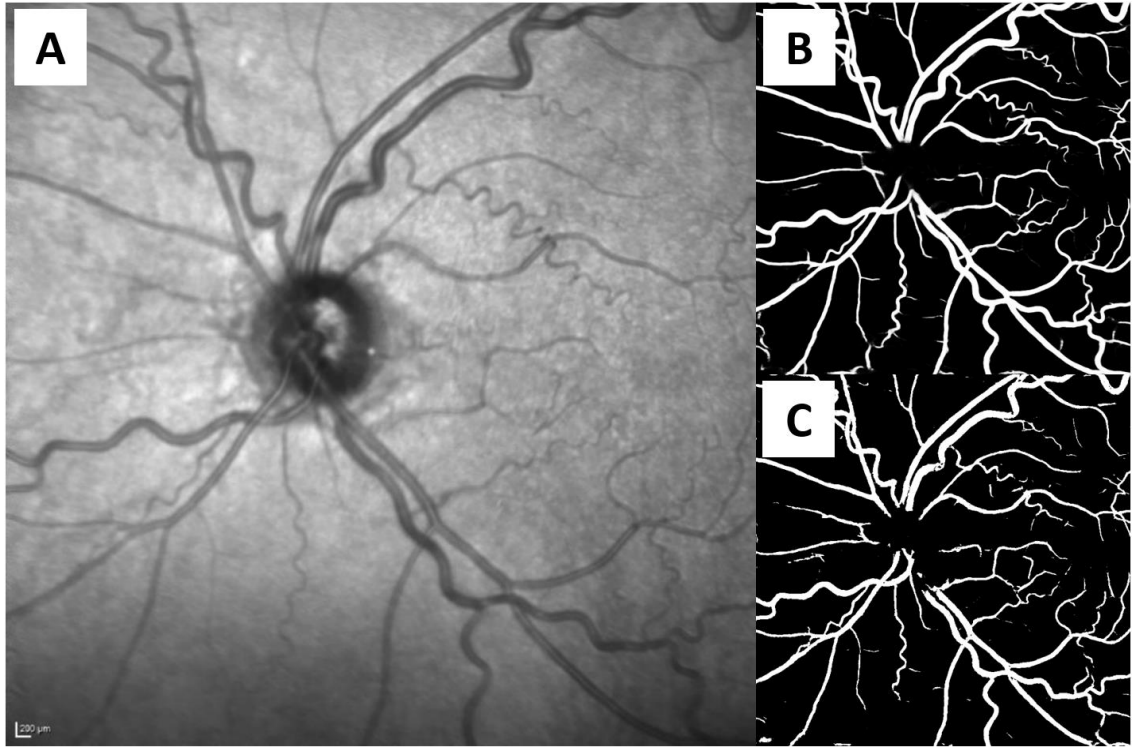

**Figure 4:** Segmentation of an SLO image (A) with two networks. Vessel map B was segmented by a network trained using data augmentation, while vessel map C was trained without data augmentation.

The current study has some limitations. Firstly, all the training images were taken with the same device model, so the network may not generalise well to other SLO devices. Additionally, on higher-resolution images (the *SPECTRALIS* can output images as either 768 × 768 pixels or 1536 × 1536 pixels), the central light reflex,[20] which appears as a bright stripe along the centre of a vessel, can affect the output vessel map, for example see the top image of Figure 3.

Despite these weaknesses, we believe the network could be a useful tool for retinal vascular analysis, for example in obtaining metrics such as fractal dimension and vessel thickness, which have proven clinically useful when extracted from colour fundus photographs. The trained network is available on GitHub, along with easy-to-follow python code.